\documentclass[aps,prl,twocolumn,amsmath,amssymb]{revtex4} 
\usepackage[english]{babel}
\usepackage{latexsym}
\usepackage{graphics}
\usepackage{subfigure}
\usepackage{epsfig}
\begin{document}

\title{Atomic quantum gases in Kagom\'e lattices}
\author{L. Santos$^{1}$, M.A. Baranov$^{1,2}$, J.I. Cirac$^{3,4}$, H.-U. Everts$^{1}$,
H. Fehrmann$^{1}$, and M. Lewenstein$^{1,4}$} 
\affiliation{(1) Institut f\"ur Theoretische Physik, Universit\"at Hannover, D-30167 Hannover,Germany\\
(2) RRC Kurchatov Institute, 123182 Moscow, Russia\\
(3) Max Planck Institut f\"ur Quantenoptik, 85748 Garching, Germany\\
(4) Institut de Ci\`encies Fot\`oniques, Barcelona 08034, Spain} 
\begin{abstract}
We demonstrate the possibility of creating and controlling an ideal and {\it trimerized} 
optical Kagom\'e lattice, 
and study the low temperature physics of various atomic gases in such lattices.
In the trimerized Kagom\'e lattice,
a Bose gas exhibits a Mott transition with fractional filling factors, 
whereas a spinless interacting Fermi gas at $2/3$ filling behaves as a quantum magnet on a triangular lattice.  
Finally, a Fermi-Fermi mixture at half filling for both components represents a frustrated quantum antiferromagnet
with a resonating-valence-bond ground state and quantum spin liquid behavior dominated by continuous spectrum of 
singlet and triplet excitations. We discuss the method of preparing and observing such quantum spin liquid 
employing molecular Bose condensates.    
\end{abstract}


\maketitle


During the last $30$ years, condensed matter physics has devoted a considerable interest 
to the issue of frustrated quantum antiferromagnets (QAF) (cf. \cite{auerbach,sachdeev}). 
In 1973 P.W. Anderson proposed \cite{anderson} the resonating-valence-bond (RVB) 
for the ground state of QAFs, where all spins are paired into singlets. 
RVB states exhibit neither the standard antiferromagnetic N\'eel order, 
nor the spin-Peierls order (for which singlet pairs are ordered in space). 
Recent extensive numerical studies have shown that the 
RVB physics characterizes the spin $1/2$ Heisenberg antiferromagnet 
on the 2D Kagom\'e lattice in 2D (see Fig. 1a) \cite{waldmann,moessner}. 
The spectrum of this system has a very peculiar structure:
the energy gap between the ground state and the lowest triplet state,
if any, is predicted to be very small (of order $J/20$, where $J$ is the
spin exchange coupling). This gap is filled with low-lying singlet states. 
Their number scales as $1.15^N$ with the number of lattice sites $N$. 
For temperatures above the triplet gap, 
the spin correlations decay rapidly in space, but have a very slow temporal behavior
$\langle s(x,0)s(x,t)\rangle \propto 1/t^{0.6}$ \cite{georges}. 

A very illuminating analytic insight into the physics of QAF in Kagom\'e lattice has been recently
 obtained by Mila and Mambrini \cite{mila}, who considered a {trimerized} Kagom\'e lattice 
(TKL, see Fig. 1b). Such a lattice is a triangular lattice of trimers 
with intra- (inter-)trimer couplings $J$ and $J'\ll J$, respectively. In Refs. \cite{mila} 
a non-trivial mean field theory has been formulated that predicts 
correctly the number, the form, and the spectrum of singlet excitations, 
which correspond to a restricted set of short-range RVB states.  
For $J'\ll J$ the theory predicts a triplet gap $(2/3)J'$. All these 
 theoretical findings do not yet have a clear experimental confirmation 
in condensed matter systems, and it is thus desirable to seek
 for other possible testing grounds. 

Such novel testing grounds could be provided by the atomic 
physics of ultracold quantum lattice gases, 
which is one of the most rapidly developing fields of AMO physics nowadays. 
Following the proposal of Jaksch {\it et al.} \cite{jaksch}, 
Greiner {\it et al.} \cite{greiner} were the first to demonstrate  
the superfluid-Mott insulator (MI) transition in a lattice Bose gas, 
predicted earlier in Ref. \cite{fischer}.
 Atomic physics and quantum optics offer in this context 
a new and very precise way of preparing, manipulating and detecting the 
system under investigation. 

In this Letter we show that using superlattices 
techniques \cite{superlattices}, 
it is possible to create a 2D optical trimerized Kagom\'e lattice, 
and to control 
in real time the degree of trimerization (i.e. the ratio of $J'/J$). The physics of
cold atomic gases in such an optical lattice is described quite generally by various versions 
of the Hubbard model, and the energies and couplings defining such models can be calculated 
using solid-state methods (tight binding approximation, 
Wannier function expansion \cite{mermin}). Such unprecedented possibility 
motivates us to  consider three kinds of quantum gases in the TKL: 
i) A Bose gas with repulsive interactions; for this case we predict the appearance 
of Mott-insulator phases with fractional fillings $\nu=1/3, 2/3$; 
ii) A spinless Fermi gas with nearest-neighbor interactions; such gas appears for instance 
naturally as a strong-interaction limit of the Bose-Fermi Hubbard model \cite{lewen}. 
In this case fermions (or, more generally composite fermions consisting of bare fermions 
coupled to bosons or bosonic holes) attain boson mediated interactions. Interestingly, at $2/3$ filling
the Fermi gas behaves as a frustrated quantum magnet on a triangular lattice with couplings 
dependent on the direction of the bonds; 
iii) Finally, we consider a Fermi-Fermi mixture at full total filling $N_1=N_2=N/2$, 
where $N_i$ is the number of fermions of each species. Such a system in the strong 
coupling limit is equivalent to the spin $1/2$ Heisenberg antiferromagnet in zero magnetic field,  
and the physics of this system is well-described by the RVB model.
We discuss in details how to prepare a system in low-lying singlet states, 
and how to detect its properties.


In the following, we consider the atoms in a 2D optical lattice 
in the $xy$ plane, being strongly confined (magnetically or optically) 
in the $z$ direction. 
In order to form  a Kagom\'e lattice one can use  blue detuned lasers,
so that the potential minima coincide with the laser intensity maxima.
A perfect triangular lattice can be easily created by  
two standing waves on the $xy$ plane, $\cos^2(\vec k_{1,2}\vec r)$, 
with $\vec k_{1,2}=k \{1/2,\pm \sqrt{3}/2 \}$,
and an additional standing wave 
$\cos^2(\vec k_3\vec r +\phi)$, with $\vec k_3=k \{0,1 \}$. 
The resulting triangles have a side of length $2\pi/\sqrt{3}k$. 
By varying $\phi$ the third standing wave 
is shifted along the $y$ axis, and, in principle, 
a Kagom\'e pattern could be realized. 

However, this procedure presents two problems. First, $3$ lasers on a plane cannot 
have a mutually orthogonal polarization, and consequently 
undesired interferences between different 
standing waves occur. This can be avoided
by randomizing the relative orientation of the polarization between different 
standing waves, or by introducing small frequency mismatches, which, however,
have to be larger than any other relevant frequencies. 
Second, due to the diffraction limit, the ratio $\xi$ between the separation 
between maxima and the half-width at half maximum (HWHM) 
is $4$. Because of that, for any $\phi$, the three potential minima forming the 
Kagom\'e triangles cannot be resolved.
This can be avoided by using superlattices. 
For instance, one may substitute each standing wave 
($i=1,2,3$) by a laser potential 
$(\cos(\vec k_i\vec r)+2\cos(\vec k_i\vec r/3))^2$. For this potential
$\xi=7.6$, and the generation of a perfect Kagom\'e lattice for $\phi=\pi/2$, and 
a modestly TKL for $5\pi/12\le\phi\le\pi/2$ is possible.  Using 
another superlattice:
$(\cos(\vec k_i\vec r)+2\cos(\vec k_i\vec r/3)+4\cos(\vec k_i\vec r/9))^2$, 
one can reach 
$\xi\sim 14$, and generate for $\phi=\pi/4$ a strongly   TKL 
(Fig.~1(b)). The necessary laser 
configuration is sketched in  Fig.~1(c). 


\begin{figure}[ht]
\begin{center}
\psfig{file=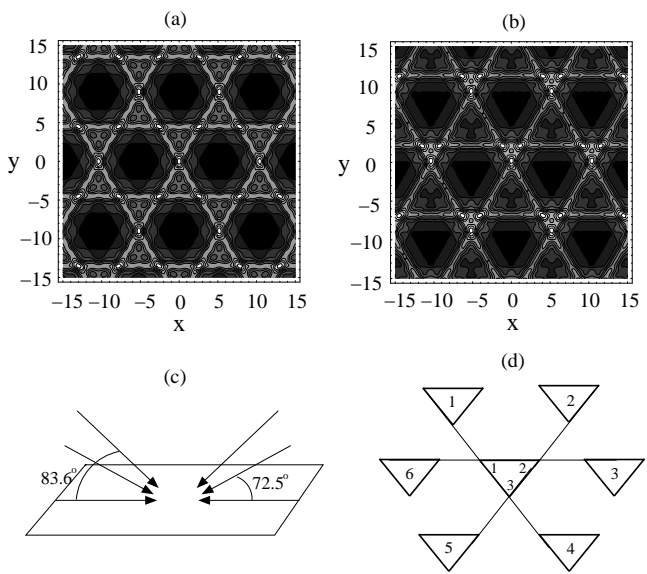,width=8.5cm}
\end{center}
\caption{(a) Ideal Kagom\'e lattice for $\phi=\pi/2$; 
(b) TKL using $\phi=\pi/4$. 
This lattice can be generated using 3 standing waves with $\pi/3$ angle 
between themselves. 
Each standing wave is generated by 3 lasers with a configuration shown in (c); 
(d) enumeration of spins in a trimer and of neighboring trimers, 
used in Eq. (3).
}
\label{fig:1}
\end{figure}


The proposed scheme allows one to realize 
the trimerized and the ideal Kagom\'e lattice, and transform one into 
the other and back by changing $\phi$. 
One is thus tempted to study the physics of ultracold gases in such lattices. 
In the following, we consider three different physical systems: a Bose gas, 
a single component Fermi gas with nearest-neighbor interactions, 
and a two-component Fermi gas.
The physics of these systems is quite generally described by a Hubbard model
\cite{jaksch}. The parameters of the model include 
intra- (inter-)trimer tunneling (hopping) $J$ ($J'$), 
on-site interaction energies $V$ for different species, and 
nearest-neighbor interactions $U$ ($U'$). These parameters can be controlled 
by modifying the laser detunings and intensities, as well as the 
configuration of the superlattices (the phase $\phi$).

To calculate the parameters $J, J',V,\ldots$ we determine first 
the Bloch functions for the problem. 
Since there are $3$ lattice sites per Bravais lattice, 
the lowest band splits actually  into 
three different bands \cite{mermin}. For each  of the bands we calculate 
the corresponding Wannier functions (localized at each one of the Kagom\'e triangles), 
and combine them then by means of an orthogonal transformation to 
construct functions localized at each lattice site. Once these functions are known, 
the tunneling and interaction coefficients are calculated as  in Ref.~\cite{jaksch}. 


We consider first the Bose gas, which is described by the corresponding Bose-Hubbard Hamiltonian
\begin{eqnarray}
&&H_{\mathrm{BH}}=-\sum_{\left\langle ij\right\rangle }
(J_{ij} b_{i}^{\dagger }b_{j}+ {\rm h.c.}) 
+\sum_{\left\langle ij\right\rangle}
U_{ij} n_{i}n_{j}\nonumber \\
&&+\sum_{i}\left[
\frac{1}{2}Vn_{i}(n_{i}-1)-\mu n_{i}\right] , 
\label{BH} 
\end{eqnarray}
where the tunneling and nearest-neighbor interaction coefficients 
$J_{ij}$ and $U_{ij}$ take here the values $J$ and $U$ for intra-, and
$J'$ and $U'$ for inter-trimer hopping,
$\mu$ is the chemical potential, $n_i=b_i^\dag b_i$,
and $b_i, b_i^\dag$ are the annihilation and creation operators for bosons at the site $i$. 
In the limit when the boson number $N_B\le N$, i.e for 
filling factors $\nu\le 1$, and for strong on-site interactions 
$V\gg J,J'$, only one particle may occupy a lattice site.  

In the strongly trimerized case ($ J', U'\ll  U< J$) the system will 
enter a trimerized Mott phase 
with the ground state corresponding to the product over (independent) trimers. 
Depending on the particular value of $\bar\mu\equiv (\mu-U)/(2J+U)$ we may have 
$0$ ($\bar\mu<-1$), $1$ ($-1\le \bar\mu<0$), $2$ ($0\le \bar\mu<1$) or $3$ ($1\le \bar\mu$) 
bosons per trimer, i.e. filling factors $\nu=0$, $1/3$, $2/3$ or $1$ boson per site.
For fractional filling, the atoms within a trimer minimize the 
energy forming a, so-called, W-state \cite{W}: 
$|W\rangle=(|001\rangle+ |010\rangle +|100\rangle)/\sqrt{3}$ for $\nu=1/3$, 
and $|W\rangle=(|110\rangle +|101\rangle +|011\rangle)/\sqrt{3}$ for $\nu=2/3$. 
It is worth noticing that $W$-states themselves have interesting applications for 
quantum information theory(c.f. \cite{applW}).

Generalizing the Landau mean-field theory
of Ref.~\cite{fischer}, 
we obtain the phase diagram in the $\bar J'\equiv J'/(2J+U)$ and $\bar\mu$ 
plane
with characteristic lobes describing the boundaries of the Mott phases, given by
$\bar J'=(|\bar\mu|-1)/2$ for $|\mu|\ge 1$, and 
$\bar J'=(3/2)|\bar\mu|(1-|\bar\mu|)/(4-|\bar\mu|)$ for  $|\mu|< 1$. 
Observations of this Mott transition require temperatures $T$ of the 
order of $ J'$, i.e. smaller 
than $ J$ and $ U$. Assuming that $ U$ is of the order of few recoil energies \cite{greiner}, 
that implies $T$ in the range of tens of nK. The results for $ J< U$ are qualitatively similar.  


The spinless Fermi gas in the trimerized Kagom\'e lattice 
is described by the Fermi-Hubbard  Hamiltonian
\begin{equation}
H_{\mathrm{FH}}=-\sum_{\left\langle ij\right\rangle }
(J_{ij} f_{i}^{\dagger }f_{j}+ {\rm h.c.}) 
+\sum_{\left\langle ij\right\rangle}
U_{ij} n_{i}n_{j}-\sum_{i}\mu n_{i}, 
\label{FH} 
\end{equation}
where as before $J_{ij}$ and $U_{ij}$ take the values $J$ and $U$ for intra-,  
and $J'$ and $U'$ for inter-trimer hopping,  
$\mu$ is the chemical potential, $n_i=f_i^\dag f_i$, 
and $f_i, f_i^\dag$ are the fermionic annihilation and creation operators.
In the following,  we enumerate the sites in each trimer 
$1$, $2$, $3$ clockwise starting from the upper left site. We 
denote the 3 different intra-trimer modes by 
 $f=(f_1+f_2 +f_3)/\sqrt{3}$ (zero momentum mode), 
and  $f_{\pm}=(f_1+z_{\pm}f_2 +z_{\pm}^2f_3)/\sqrt{3}$ (left and right 
chirality modes), 
where $z_{\pm}=\exp(\pm 2\pi i/3)$. The intra-trimer Hamiltonian  acquires the form
$
H_{intra}= -3 J f^\dag f +  U[(\bar n-\tilde\mu)^2-\tilde\mu^2]/2,
$
where $\bar n$ is the total fermion number in the trimer, 
whereas $\tilde \mu= (\mu-J+U/2)/U$.  
In the strongly trimerized limit, we have (depending on the value of 
$\tilde\mu$): 
$0$ ($\tilde\mu<1/2-3J/U$), $1$ ($1/2-3J/U<\tilde\mu<3/2$), $2$ ($3/2<\tilde\mu<5/2$) or $3$ 
($\tilde\mu>5/2$) fermions per trimer.   
Obviously, the cases with $0$ and $3$ fermions per trimer are not interesting. 
In the case of $1$ fermion per trimer or less (filling factor $0\le \nu\le 1/3$), 
at low temperatures ($T<J$)
the fermions will preferably occupy the lowest energy zero momentum mode $f$
forming a gas of $f$-fermions on a triangular lattice.
This gas will have a tunneling rate $J'/3$, 
and  a coupling constant for nearest-neighbor interactions $U'/9$. Depending on the sign of $U'$ 
we expect here similar behavior to that discussed in Ref.~\cite{lewen}, i.e. the appearance of 
a superfluid phase or fermionic domains for $U'<0$, and  Fermi liquid, or density wave phases for $U'>0$.

The situation is more complex in the case of $2$ fermions per 
trimer, since  the 
second fermion may then have left, or right chirality, while 
the first fermion occupies the $f$ state with certainty. The 
system becomes equivalent to a quantum magnet with interactions
 that depend on the bond directions, described by the Hamiltonian:
\begin{equation}
H_{int}= \frac{2U'}{9}\sum_i \sum_{j=1}^6 s_i(\phi_{i\to j})  
s_j(\tilde\phi_{j\to i}),
\label{hferint1}
\end{equation} 
where the nearest neighbors are enumerated clockwise as shown in Fig~\ref{fig:1}(d). 
In Eq.~(\ref{hferint1}) 
we have $s(\phi)=\cos(\phi)s_x + \sin(\phi)s_y$, where 
$s_x=(f_+^\dag f_- + f_-^\dag f_+)/2$ and 
$s_y=-i(f_+^\dag f_- - f_-^\dag f_+)/2$. The angles $\phi$ are 
$\phi_{i\to 1}=\phi_{i\to 6}=0$, 
$\phi_{i\to 2}=\phi_{i\to 3}=2\pi/3$, $\phi_{i\to 4}=\phi_{i\to 5}=-2\pi/3$, 
$\tilde\phi_{1\to i}=\tilde\phi_{2\to i}=-2\pi/3$, $\tilde\phi_{3\to i}=
\tilde\phi_{4\to i}=0$, and 
$\tilde\phi_{5\to i}=\tilde\phi_{6\to i}=2\pi/3$.

The couplings between the left and right modes  are effectively 
ferromagnetic for $U'>0$, 
and antiferromagnetic otherwise. Inter-trimer hopping $J'$ contributes first in order $(J')^2/U$,
and is neglected here.
For $U'>0$ the spins align ferromagnetically in the $xy$-plane in the classical ground state. 
Quantum fluctuations around this state are represented in the semiclassical approximation 
by gapless magnons (cf.~\cite{auerbach}). For $U'<0$ the classical approximation yields a 
N\'eel-type ground state with planar order 
in which on every triangle there is exactly one spin directed along the 
$\phi=0, 2\pi/3, -2\pi/3$ directions. Due to the anisotropy of the Hamiltonian 
there is no continuous rotational invariance in spin space. Therefore this planar 
ground state is unique (up to discrete $2\pi/3$-rotations), and hence the semiclassical 
excitation spectrum in this case is gapped with a gap of order $|U'|$.

The observation of the quantum phases in this model requires $U'<J,U$, and $T<U'$. 
If the fermionic interactions are due to dipolar forces \cite{dipole}, then $U$ may be of order 
of few recoils, and $T$ in the range of $10$-$100$ nK. If fermionic interactions 
are due to hopping induced effects in a Bose-Fermi mixture (as in Ref.~\cite{lewen}), 
then $T$ in the range of $10$ nK will be  necessary.


Finally, we consider a Fermi-Fermi mixture with half filling 
for each species, i.e. a spin $1/2$ Hubbard model
\begin{equation}
H_{FF}=-\sum_{\left\langle ij\right\rangle} J_{ij}
(f_{i}^{\dagger }f_{j}+ \tilde f_{i}^{\dagger }\tilde f_{j} + {\rm h.c.}) +
\sum_i V  n_{i}\tilde n_{i}, 
\label{FFH}
\end{equation}
where the operators $f_i$ and $f_i^\dag$ ($\tilde f_i$ and $\tilde f_i^\dag$) are the creation and annihilation 
operators for the two components, $n_i=f_i^\dag f_i$ ($\tilde n_i=\tilde f_i^\dag \tilde f_i$), 
and, as above, $J_{ij}=J_0$ ($J'_0$) for intra- (inter-) trimer hopping. In the strong coupling limit, $J_0,J'_0\ll V$ 
($t-J$ model) \cite{auerbach}, $H_{FF}$ reduces to the Heisenberg
 antiferromagnet
\begin{equation}
H=J\sum_{\langle i,j\rangle_{intra}}\vec s_i \cdot \vec s_j+
\bar J'\sum_{\langle i,j\rangle_{inter}}\vec s_i \cdot \vec s_j,
\end{equation}
where $J=4J_0^2/V$, and $J'=4{J'}_0^2/V$,	and 
$\vec s=(s_x,s_y,s_z)$, with $n-\tilde{n}=2s_z$, 
$f^\dag \tilde f=s_x+is_y$, and $\tilde f^\dag f=s_x-is_y$.

In the strongly trimerized limit \cite{mila}, the total spin in the trimer takes the minimal value, i.e. $1/2$, 
and there are four degenerate states having $s_z=\pm 1/2$ and  
left or right  chirality.  
The spectrum of the system in the singlet 
sector consists of a narrow band of low energy states 
of the width of order $J'$, 
 separated from the higher singlet (triplet) 
bands by a gap of order {\bf $3J/4$ ($2J'/3$)}.

Assuming that we can prepare the system in a singlet state at $J'<T<J$, 
then the density of states of the low lying singlet levels can be obtained 
by repeated measurements of the system energy. 
The latter can be achieved by simply releasing the lattice, so that after 
taking care of the zero point energy, 
all of the interaction energy is transformed into kinetic energy. 
In a similar way we can measure the mean value and the distribution 
of any nearest-neighbor two-spin correlation functions. To this aim one has to  
 apply at the moment of the trap release a chosen nearest neighbor two-spin 
Hamiltonian and keep it acting during the cloud expansion 
(for details see \cite{ciracjuanjo}).  In a similar manner we can measure the 
spectrum of triplet excitation, by exciting a triplet state, 
which can be done by flipping one spin 
using a combination of superlattice methods and laser 
excitation \cite{blochpri}. The measurement of the singlet-triplet gap
requires a resolution better than $J'$.

A similar type of measurements can be 
performed in the ideal Kagom\'e lattice, when $J=J'$. In this case, 
the singlet-triplet gap is filled with singlet excitations \cite{waldmann}. 
By varying $\phi$, one can 
transform adiabatically from strongly trimerized to ideal Kagom\'e, for which 
the final value of $J$ will be smaller than the initial $J$, but larger than the initial $J'$. 
In that case, the system should remain within the lowest set of $1.15^N$ states
that originally formed the lowest singlet band. The singlet-triplet 
gap, if any,  is estimated to be $\le J/20$, and should be measurable 
using the methods described above.

A possible way to  prepare  a singlet state in the TKL 
with $T<3J/4$ could employ  the recently obtained 
Bose-Einstein condensates of molecules consisting of two fermionic 
atoms \cite{jin} at temperatures of 
the order of $10$ nK. Such BECs should be loaded onto an ideal and 
weak Kagom\'e lattice. 
Note that the molecules formed after sweeping 
across a Feshbach resonance, are in a singlet state of the 
pseudo-spin $\vec s$.
This can easily be seen, because the two fermions enter the resonance 
from the $s$-wave scattering channel (i.e. in the symmetric state with 
respect to the spatial coordinates), and thus are  
 in a singlet state of the pseudo-spin (i.e. antisymmetric state with 
respect to exchange of electronic and nuclear spins). 
Since the interaction leading to the 
spin flipping at the Feshbach resonance \cite{Timmermans} 
is symmetric under the simultaneous interchange of both 
nuclear and electronic spin, then the formed molecule 
remains in a pseudo-spin singlet. The typical size of the 
molecule is of the order of the $s$-wave scattering length $a$, 
and thus can be modified at the resonance \cite{Petrov},
being chosen comparable to the lattice wavelength. 
Growing the lattice breaks the molecule into two separate fermionic 
atoms in neighboring sites  in the singlet pseudo-spin state. 
In this way, a singlet covering of the Kagom\'e lattice may be achieved,
allowing for the direct generation of a RVB state \cite{anderson}.

Summarizing, we have shown that by employing currently available 
superlattice techniques it is possible to create in a controlled way 
a trimerized Kagom\'e lattice. An ultracold Bose gas 
in such lattice exhibits novel Mott insulator phases with fractional filling 
$\nu=1/3,2/3$. A single-component Fermi  gas with nearest neighbor 
interactions for $\nu=1/3$ behaves  as a Fermi gas in the underlying 
triangular lattice, whereas for $\nu=2/3$ it becomes a non-standard 
ferro- or antiferromagnet in
the  triangular lattice with direction dependent bonds. 
Finally, for 
a Fermi-Fermi mixture in a Kagom\'e lattice, which is described by 
 an antiferromagnetic Heisenberg model, 
we have shown the possibility to measure  the 
spectral properties of the latter system. This opens 
the way of analyzing in a completely novel and fascinating setting one of the 
paradigmatic problems of condensed-matter physics, 
the physics of random valence bond frustrated antiferromagnets.

We acknowledge discussions with M. Greiner, R. Grimm, and P. Julienne, and 
support from the Deutsche
Forschungsgemeinschaft (SFB 407 and SPP1116), the RTN Cold Quantum Gases, 
ESF Programme BEC2000+, the Russian Foundation for Basic Research,
and the Alexander von Humboldt Foundation.


\begin{references}

\bibitem{auerbach}  A. Auerbach, \textit{Interacting Electrons 
and Quantum Magnetism}, (Springer,New York, 1994). 
\bibitem{sachdeev} S. Sachdev, \textit{Quantum Phase Transitions},
(Cambridge University Press, Cambridge, 1999).
\bibitem{anderson} P.W. Anderson, Mater. Res. Bull. {\bf 8}, 153 (1973). 
\bibitem{waldmann} C. Waldtmann {\it et al.}, Eur. Phys. J B{\bf 2}, 501 (1998);
P. Lecheminant {\it et al.}, Phys. Rev. B{\bf 56}, 2521 (1997).
\bibitem{moessner} R. Moessner and S.L. Sondhi,  Phys. Rev. Lett. {\bf 86}, 1881 (2001); R. Moessner, S.L. Sondhi, and E. Fradkin, Phys. Rev. B{\bf 65}, 024504 (2001). 
\bibitem{georges}  A. Georges, R. Siddharthan, and S. Florens, Phys. Rev. Lett. {\bf 87}, 277203
(2001). 
\bibitem{mila} F. Mila, Phys. Rev. Lett. \textbf{81}, 2356
(1998); F. Mila and M. Mambrini, Eur. Phys. J. B{\bf 17}, 651 (2000). 
\bibitem{jaksch}  D. Jaksch \textit{et al.}, Phys. Rev. Lett. \textbf{81}, 3108
(1998).
\bibitem{greiner}  M. Greiner \textit{et al.}, Nature \textbf{415}, 39 (2002).
\bibitem{fischer} M.P.A. Fisher, P.B. Weichman, G. Grinstein, and D.S. Fisher, 
Phys. Rev. B {\bf 40}, 546 (1989).
\bibitem{superlattices} G. Grynberg {\it et al.}, Phys. Rev. Lett. {\bf 70}, 2249 (1993);
L. Guidoni and P. Verkerk, Phys. Rev. A{\bf 57}, R1501 (1998); 
P. Rabl et al. cond-mat/0304026.
\bibitem{mermin} N.W. Ashford and D.N. Mermin, {\sl Solid State Physics},
(Holt, Reinhart and Winston, Philadelphia, 1976).
\bibitem{lewen} M. Lewenstein, L. Santos, M. Baranov and H. Fehrmann, cond-ma
t/0306180, in print in Phys. Rev. Lett. (2004).
\bibitem{dipole}
K. G\'oral, L. Santos, and M. Lewenstein, Phys. Rev. Lett. {\bf 88}, 170406 (2002). 

\bibitem{W} W. D\"ur, G. Vidal, and J. I. Cirac, Phys. Rev. A{\bf 62}, 062314 (2000).

\bibitem{applW} J. Joo, Y.J. Park, S. Oh, and J. Kim, New J. Phys. {\bf 5}, Art. No. 136 (2003).

\bibitem{ciracjuanjo} J.J. Garcia-Ripoll {\it et al.} (in preparation).

\bibitem{blochpri} I. Bloch (unpublished). 


\bibitem{jin}  S. Jochin {\it et al.}, Science Express {\bf 302}, 2101 (2003); 
M. Greiner, C.A. Regal, and  D.S. Jin, Nature {\bf 426}, 537 (2003); 
M. W. Zwierlein  {\it et al.}, Phys. Rev. Lett. {\bf 91}, 250401 (2003).

\bibitem{Timmermans} E. Timmermans {\it et al.}, 
Phys. Rep. {\bf 315}, 199 (1999).

\bibitem{Petrov} D.S. Petrov, C. Salomon, and G.V. Shlyapnikov, 
cond-mat/0309010.
\end{references}
\end{document}